\documentclass[usenatbib]{mn2e}

\usepackage{epsfig}

\begin{document}


\title[Modelling realistic horizontal branches]{Modelling realistic horizontal 
branch morphologies and their impact on spectroscopic ages of unresolved stellar systems}
\author[Susan M. Percival and Maurizio Salaris]{Susan M. Percival$^{1}
$\thanks{E-mail:smp@astro.livjm.ac.uk ; ms@astro.livjm.ac.uk} 
and Maurizio Salaris$^{1}$\footnotemark[1]\\
$^{1}$Astrophysics Research Institute, Liverpool John Moores University, Twelve Quays House, Egerton Wharf, Birkenhead CH41 1LD, UK}


\date{}

\pagerange{\pageref{firstpage}--\pageref{lastpage}} \pubyear{2002}

\maketitle

\label{firstpage}

\begin{abstract}
The presence of an extended blue horizontal branch (HB) in a stellar 
population is known to affect the age inferred from 
spectral fitting to stellar population synthesis models.  This is due to 
the hot blue component which increases the strength of the Balmer lines 
and can make an old population look spuriously young.  However, most 
population synthesis models still rely on theoretical isochrones which 
do not include realistic modelling of extended HBs.  
In this work, we create detailed models for a range of old simple stellar 
populations (SSPs), with metallicities ranging from [Fe/H]=$-1.3$ to 
solar, to create a variety of realistic HB morphologies, from extended 
red clumps, to extreme blue HBs.
We achieve this by utilising stellar tracks from the BaSTI database and  
implementing a different mass loss prescription for each SSP created. 
This includes setting an average mass and a Gaussian spread in masses of 
individual stars coming on to the zero age HB for each model, 
and hence resulting in different HB morphologies.  
We find that, for each metallicity, there is some HB morphology which 
maximises H$\beta$, making an underlying 14~Gyr population look 
$\sim5-6$~Gyr old for the low and intermediate metallicity cases, and 
as young as 2~Gyr in the case of the solar metallicity SSP.
We explore whether there are any spectral indices capable of breaking the
degeneracy between an old SSP with extended blue HB and a truly young
or intermediate age SSP, and find that the Ca{\sc ii} index of \citet{rose84}
and the strength of the Mg{\sc ii} doublet at 2800\AA\ are promising candidates,
in combination with H$\beta$ and other metallicity indicators such as Mg$b$
and Fe5406.
We also run Monte Carlo simulations to investigate the level of statistical 
fluctuations in the spectra of typical stellar clusters.  We find that 
fluctuations in spectral indices are significant even for average to 
large globular clusters, and that various spectral indices are affected in 
different ways, which has implications for full-spectrum fitting methods.  
Hence we urge caution if these types of stellar clusters are to be used 
as empirical calibrating objects for various aspects of SPS models.
\end{abstract}


\section{Introduction}
\label{sec:intro}

In recent years, stellar population synthesis (SPS) models have rapidly 
become a fundamental tool in the study of both Galactic and extragalactic 
stellar populations (see \citealt{basti4,galev09,coelho07,starb99,maraston05,bc03} for some recent examples).  
Despite the increasing sophistication of the underlying
stellar models used in SPS and ever-growing libraries of spectra
(both empirical and synthetic) available to modellers, there are still
areas in which models, and methods for fitting them to observational data, 
can be significantly improved and expanded.  Some of
the current shortcomings are due to the need to adopt simplifying 
assumptions about the stellar population in question in order to make some 
problems tractable, e.g. the fitting of simple stellar population (SSP) 
models to galaxies, assuming that they can be represented by a single age, 
single metallicity model.  

A key area in which simplifying assumptions are implicitly made is the
morphology of the horizontal branch (HB) for individual SSPs -- this is 
largely due to the limitations of theoretical isochrones, which are used 
as the basis of all SPS models (except those of Maraston and co-workers, 
who use a fuel-consumption based method -- see \citealt{maraston05} and 
references therein).  However it is well known that the morphology of the HB 
can impact strongly on the integrated light of stellar populations 
especially if an extended blue component is present, potentially 
affecting both colours 
and line indices, and hence impacting on inferred ages
for these systems (see e.g. \citealt{ocvirk,conroy09,schi04,lee2002,lee2000}).
Hence it is vital to assess the ways which these `simplified' HBs inherent 
in theoretical isochrones impact on SPS models, and whether a more 
realistic and detailed treatment of the HB is necessary, or indeed practical.  
This is the purpose of the work presented here.

For unresolved stellar populations our knowledge of their ages and elemental
abundances is largely derived through the fitting of diagnostic spectral
indices, hence the focus of our work is to investigate the effect of 
a detailed treatment of the HB on key spectral indices, such as H$\beta$.  
However, it is important
to realise that our results will also impact on analyses which use 
full-spectrum fitting methods, as will be discussed in Section~\ref{sec:disc}.

In real stellar systems the horizontal extension of the 
HB is governed by stochastic mass loss in stars approaching the tip of the 
red giant branch (TRGB).  All the stars in a particular SSP leave the 
TRGB with the same core mass 
but individual stars have different masses remaining in the outer layers, 
which determine the position of the star along the zero-age horizontal 
branch (ZAHB).  In stellar models this mass loss is parameterised by the 
mass loss parameter, $\eta$, which is assigned some fixed value calculated
according to the Reimers mass loss relation \citep{reimers}.  
Hence in theoretical isochrones the HB comprises a single mass point, with
no extension.

It is worth remembering here that an isochrone consists of a series 
of evolutionary points (EPs) which define the locus of points for an SSP.
Whilst each EP defines the appropriate stellar parameters for any star
located at that point, in terms of mass, effective temperature and 
luminosity, they do not represent individual stars themselves.  
In order to create an SSP, an isochrone is ``populated'' according to 
some initial mass function (IMF) which effectively gives the appropriate 
weighting to each EP.  In most population synthesis work the isochrone 
is treated as an analytical function for this purpose, so that all EPs
along the isochrone are smoothly populated and the ZAHB remains as a 
single mass point.  In order to create a model with an 
extended HB, effectively incorporating a spread in mass loss, the 
isochrone (or, at least, the HB portion) must be populated with a 
discrete number of stars, so that each star in the HB phase can be assigned 
a specific mass from within some range of masses  -- this requires 
interpolation between individual core-helium burning stellar tracks.  

Creating an integrated spectrum for these extended HB models is a 
computationally expensive procedure since the time taken is proportional to 
the number of points in the simulation (for details on how integrated
spectra for SSPs are produced the interested reader is referred to 
\citealt{basti4}, hereafter P09).  
For the analytical case, the number of points is just 
the number of EPs along the isochrone, which varies between a few 
hundred and around two thousand depending on which isochrone set is used.  
To populate an isochrone with individual stars, one has to consider the 
problem of statistical fluctuations which are likely to arise as a result of
low numbers of stars (and hence poor sampling) in the later stages 
of evolution, including the HB phase.  This is of particular
relevance in the case of an SSP with an extended blue HB, since these 
hot blue HB stars give rise to strong Balmer lines,
which are generally used as the primary age indicators for stellar 
populations.  This problem can only be 
overcome by using large numbers of points in the simulation, potentially
up to $\sim10^6$.  In fact, part of the work presented here is to explore 
how large these statistical fluctuations can be in real stellar clusters, 
and how many input stars are needed in the models to avoid significant 
uncertainties in ages and metallicities derived from the final integrated 
SSP spectra.

Another problem in creating models with extended HBs is that there is 
still no theory which predicts mass loss rates from stellar parameters 
and so the value of $\eta$ is chosen arbitrarily, usually to reflect the
typical HBs seen in Galactic globular clusters.  This is complicated by the
fact that clusters apparently with the same age and chemical composition
can have different HB morphologies.  However, in principle $\eta$ can take 
any value and so producing a database of SPS models to cover all
possibilities is not feasible.  

The pioneering work of \citet{lee2000} (hereafter L2000) highlighted the 
importance of including realistic HBs in stellar population work.  L2000
created models for 15 SSPs using a single value for the mass loss parameter 
for all their models, which was chosen to replicate the range of HBs seen 
in Galactic globular clusters.  They included a Gaussian spread in the HB mass
distribution of $\sigma_{M}$=0.02$M_\odot$ to simulate the observed
extension of the HB.  This work graphically demonstrated the systematic
variation of HB morphology with age and metallicity (see their Figure 5), 
showing that, at fixed $\eta$, HBs generally become bluer as metallicity 
decreases and as age increases.  
L2000 note that the strength of the H$\beta$ line in the 
integrated population does not increase monotonically as metallicity 
decreases at a fixed age, but 
peaks at some intermediate metallicity and then falls again as 
metallicity continues to decrease.  This is because H$\beta$ reaches a 
maximum strength in stellar spectra when the effective temperature, 
$T_{eff}$, is around 9500K and so H$\beta$ is maximised if the distribution 
of stars along the HB centres around this value.  In these cases 
the contribution to H$\beta$ from the HB completely dominates over
the contribution from the main-sequence turnoff and makes the population
look spuriously young.

It is apparent from the L2000 figures, but not specifically noted by them, 
that the actual extension (i.e. from red to blue) of the HB in the 
colour-magnitude diagrams is very different for the various SSPs, even though
the value of $\eta$ and $\sigma_{M}$ are the same. 
This is because the stellar effective temperature in this phase is 
extremely sensitive to very small changes in envelope mass and so 
the resulting HB morphologies can be very different, even when the same 
mass loss prescription is used.

In this paper we use similar techniques to L2000 to model extended HBs, 
populating them with discrete numbers of stars by interpolating between 
core-helium burning stellar tracks and using a Gaussian spread in the 
mass loss distribution, and we extend that work in 3 key ways.  Firstly
we explore the effects of varying $\eta$ at fixed age and metallicity, which
enables us to model high mass loss in metal rich systems to see whether
we can mimic a system such as M32, which has approximately solar 
metallicity, but an extended blue HB component.  We are also able to 
create SSPs at different metallicities with bimodal HBs, similar to those 
observed in several Galactic globular clusters.  
Secondly, we create several models which we populate with numbers of stars 
representative of typical stellar clusters, and use Monte Carlo techniques 
to explore the impact of statistical fluctuations on measured 
diagnostic line indices, such as H$\beta$.  Thirdly, we create full
integrated high resolution spectra for all our models and simulations 
which enables us to assess the behaviour of all diagnostic line indices
(within a wavelength range 2500\AA\ to $\sim$6000\AA), 
as well as the continuum flux.  We also investigate whether there are 
any diagnostic indices capable of breaking the degeneracy between an 
old SSP with extended blue HB (hence strong H$\beta$) and a truly young 
or intermediate age SSP.

\section{Modelling SSPs with extended Horizontal Branches}
\subsection {Method}
\label{sec:modelmethod}

\begin{figure}
\includegraphics[width=80mm,angle=0]{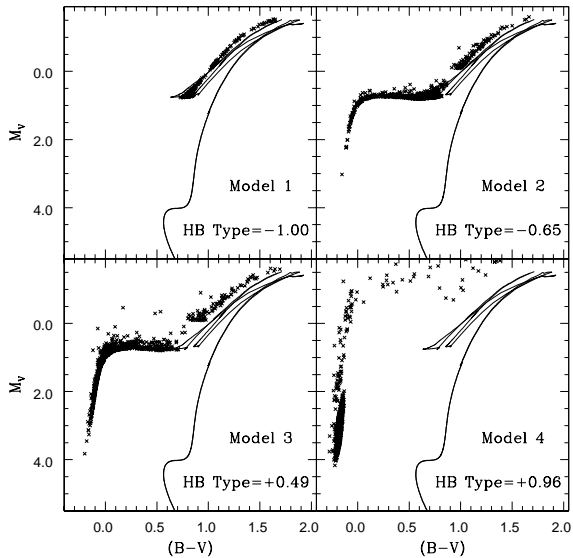}
\caption{4 models from the $\alpha$-enhanced, [Fe/H]=$-0.70$ set. For clarity, 
only 500 points are plotted for each HB ($\sim$10000 were used in each 
simulation).  The 14~Gyr isochrone used for the underlying population up to
the TRGB is also plotted.  For reference purposes, the later evolutionary 
stages of the isochone are also shown for the 2 `standard' cases in the BaSTI 
database, i.e. with fixed $\eta$=0.2 and $\eta$=0.4.  
Horizontal branch morphology
is parameterised by HB type=$(B-R)/(B+V+R)$ -- see text for details.}
\label{fig:4mods}
\end{figure}

\begin{table}
 \centering
 \begin{minipage}{80mm}
  \caption{Parameters used for $\alpha$-enhanced HB models.}
  \begin{tabular}{cccc}
  \hline
   Model & [Fe/H] & $<M>$ ($M_{\odot}$) & $\sigma_{M}$ \\
\hline            
   1 &  $-$1.31 & 0.70 & 0.02  \\
   2 &  $-$1.31 & 0.64 & 0.03  \\
   3 &  $-$1.31 & 0.58 & 0.03  \\
   4 &  $-$1.31 & 0.51 & 0.005 \\
\hline            
   1 &  $-$0.70 & 0.71 & 0.02  \\
   2 &  $-$0.70 & 0.60 & 0.03  \\
   3 &  $-$0.70 & 0.55 & 0.02  \\
   4 &  $-$0.70 & 0.495 & 0.005  \\
\hline
\end{tabular}
\end{minipage} 
\label{tab:aepar}
\end{table}

\begin{table}
 \centering
 \begin{minipage}{80mm}
  \caption{Parameters used for scaled-solar HB models.}
  \begin{tabular}{cccc}
  \hline
   Model & [Fe/H] & $<M>$ ($M_{\odot}$) & $\sigma_{M}$ \\
\hline            
   1 &  $-$1.27 & 0.72 & 0.02   \\
   2 &  $-$1.27 & 0.65 & 0.02   \\
   3 &  $-$1.27 & 0.58 & 0.02   \\
   4 &  $-$1.27 & 0.53 & 0.01   \\
\hline            
   1 &  +0.06 & 0.70 & 0.03   \\
   2 &  +0.06 & 0.63 & 0.03   \\
   3 &  +0.06 & 0.56 & 0.02   \\
   4 &  +0.06 & 0.51 & 0.01   \\
\hline
\end{tabular}
\end{minipage} 
\label{tab:sspar}
\end{table}

As a starting point, we created 16 SSPs with extended HBs, in 4 sets of 4 
models each: these comprise 2 sets with scaled-solar abundance ratios, at 
[Fe/H]$=-1.27$, $+0.06$ (solar metallicity) and 2 sets with $\alpha$-enhanced 
ratios, at [Fe/H]$=-1.31$, $-0.70$.  In each case the underlying population 
was created from an appropriate metallicity 14~Gyr isochrone from the BaSTI 
database\footnote{Available at http://albione.oa-teramo.inaf.it/} 
\citep{basti1,basti2}.  For each model the isochrone 
was populated analytically up to the TRGB and the integrated spectrum for 
this portion of the isochrone was created using high resolution spectra
from \citet{munari}, as described in P09.  
The HB was treated separately and populated with a large number of individual 
points ($>10000$, in order to avoid the problem of statistical fluctuations) 
with some mean mass (corresponding to some value of
$\eta$), and with a Gaussian spread in the mass distribution, $\sigma_{M}$, 
by interpolating between core-helium burning stellar tracks.  
Spectra were assigned to each point in the HB simulation as usual, 
i.e. matching $T_{eff}$, log $g$, and [Fe/H], and these were then 
summed together.  The summed HB spectrum was then scaled appropriately before 
being added to the spectrum of the underlying population to create the 
final integrated spectrum for each model.  

Equivalent widths of various diagnostic line indices were measured
on the resulting integrated spectra, as described in P09,
so that comparisons could be made with the BaSTI SPS database.  
Briefly, the line indices discussed here are from the Lick/IDS bandpasses 
tabulated in \citet{trager98}, unless otherwise stated.  The quoted 
line strengths were obtained using the LECTOR programme by 
A. Vazdekis\footnote{See http://www.iac.es/galeria/vazdekis/index.html} and  
are those directly measured on the spectra, i.e. they are {\it not} 
transformed onto the Lick system.  We focus
in particular on the H$\beta$ index in the following section since it
is used primarily as an age indicator for SSPs.

In each set of models, the 
4 cases were created to approximately match the extremes of HBs seen in 
Galactic globular clusters, from a purely red, but extended HB (i.e.
a red clump -- HB model 1) to an extremely blue HB with extended blue tail
(HB model 4), plus 2 intermediate cases (HB models 2 and 3). 
We also combined spectra from the extreme red and 
extreme blue cases to create models with a bi-modal HB.  Parameters used 
for each set of models, in terms of metallicity, mean mass $<M>$, and
$\sigma_{M}$, are tabulated in Tables 1 and 2, respectively, for the 
$\alpha$-enhanced and scaled-solar sets.

Figure~\ref{fig:4mods} shows the V/(B-V) colour magnitude diagrams (CMDs) 
for the $\alpha$-enhanced, [Fe/H]$=-0.70$ set of 4 models, which illustrates 
the typical HB morphologies for each set.  The HB morphological type 
$(B-R)/(B+V+R)$, as defined in \citet{lee94}, was calculated using the 
boundaries of the instability strip defined in \citet{basti3}, and 
is notated on each plot.
The appropriate isochrone (used to create the underlying population 
up to the TRGB) is also plotted for the 2 `standard' cases in the BaSTI 
database, i.e. with fixed $\eta$ of 0.2 and 0.4.  At this intermediate
metallicity the ZAHBs for the $\eta$=0.2 and $\eta$=0.4 cases have similar 
effective temperature, hence there is only a small difference in 
their location in the CMD.  The $\eta$=0.4 case always produces a hotter
(i.e. bluer) ZAHB then the $\eta$=0.2 one, but for metallicities of 
[Fe/H]$\sim-0.70$ and greater, this still produces a ZAHB
which is cooler (i.e. redder) than the turn off (TO), even at old ages.
However, for metallicities lower than this, at old ages, the $\eta$=0.4 
case produces a 
ZAHB which is very different from the $\eta$=0.2 case, and is much hotter 
(bluer) than the TO.  The difference between the 2 cases, and the impact on
derived ages, is discussed in Section~\ref{sec:hbresults}.

\subsection{Results}
\label{sec:hbresults}

\begin{figure}
\includegraphics[width=80mm,angle=0]{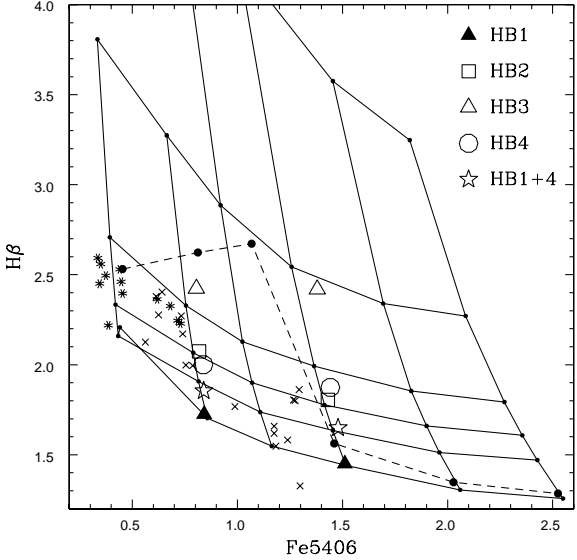}
\caption{H$\beta$ vs. Fe5406 $\alpha$-enhanced grid with results 
for the 2 sets of $\alpha$-enhanced models overplotted.  
The underlying grid is from the BaSTI SPS database, with fixed $\eta=0.2$, for 
SSP ages 1.25, 3, 6, 8, 10 and 14~Gyr (from top to bottom) and 
[Fe/H]=$-$1.84, $-$1.31, $-$1.01, $-$0.70, $-$0.29, +0.05 (increasing from 
left to right).  The dashed line is for the 14~Gyr BaSTI SSPs with fixed
$\eta=0.4$. Crosses and asterisks are observational data for Galactic GCs from 
\citet{schi05} -- the asterisks denote clusters with HB type $>$ 0.8 
(i.e. predominantly blue).}
\label{fig:aegrid}
\end{figure}

\begin{figure}
\includegraphics[width=80mm,angle=0]{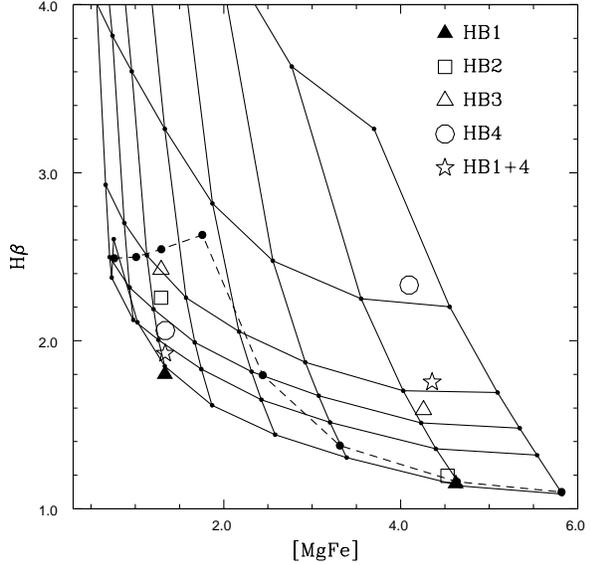}
\caption{H$\beta$ vs. [MgFe] scaled-solar grid with results for the
2 sets of scaled-solar models overplotted.  The underlying grid is as
in Figure~\ref{fig:aegrid} but for 
[Fe/H]=$-$1.79, $-$1.49, $-$1.27, $-$0.96, $-$0.66, $-$0.35, +0.06, +0.40.}
\label{fig:ssgrid}
\end{figure}

Figure~\ref{fig:aegrid} shows results from the $\alpha$-enhanced HB models 
in the H$\beta$ v. Fe5406 diagnostic grid, i.e. where H$\beta$ is the
age indicator and Fe5406 traces iron abundance, [Fe/H].  
Similarly, Figure~\ref{fig:ssgrid} shows results from the scaled-solar 
HB models in the H$\beta$ vs. [MgFe] grid, where [MgFe] is a tracer of 
total metallicity, $Z$ (see P09).  In these figures, HB1 refers to the 
red clump only HB model, HB4 refers to the extended blue tail case and 
HB2 and HB3 are the intermediate cases.  The underlying
models from which these grids are constructed are the BaSTI $\eta$=0.2
models (solid lines), with the $\eta$=0.4 models at 14~Gyr only overlaid
(dashed lines), described in P09.

Before discussing the results of the extended HB simulations it is important
to note the behaviour of H$\beta$ in the fixed $\eta$=0.4 models, at old ages, 
compared to the $\eta$=0.2 ones.  
Figures~\ref{fig:aegrid} and \ref{fig:ssgrid} clearly demonstrate that at
metallicities above [Fe/H]$\sim-0.7$, the H$\beta$ line has similar strength
for the 2 cases, indicating that their HBs have similar average temperatures
and both appear as red clumps in a CMD.  However between [Fe/H]$=-0.7$
and [Fe/H]$=-1.0$, the H$\beta$ line suddenly becomes much stronger for 
the $\eta$=0.4 case, which is an indication that the HB stars are much hotter
than for the $\eta$=0.2 case, and hence the HB would appear much bluer
than the turn off in a CMD.  For the $\eta$=0.4 models, H$\beta$ peaks at
a metallicity of around [Fe/H]$=-1.0$ and then decreases again as
the metallicity continues to decrease.  This is the same behaviour noted 
by L2000 and happens because the average temperature of the HB stars gets
increasingly hot as the metallicity decreases until it is significantly 
hotter than 9500K (where the H$\beta$ strength peaks).

In all 4 model sets, an extended red only HB (model HB1) only negligibly 
affects H$\beta$ compared to the `standard' BaSTI SSP model with single mass 
ZAHB and $\eta$=0.2, hence the inferred age from these diagnostic grids is
the correct one for the underlying stellar population, i.e. 14~Gyrs.  
For the 3 sets of models with sub-solar metallicity (i.e. the 2 
$\alpha$-enhanced sets and one of the scaled-solar sets), H$\beta$ reaches 
a peak for one of the intermediate morphology cases (model HB3), and is 
similar to the fixed $\eta$=0.4 case, whilst for the scaled-solar 
solar metallicity set, H$\beta$ peaks for model HB4, i.e. the extended blue
tail case.  For the lower metallicity models, at [Fe/H]$\sim -1.3$, 
the peak in H$\beta$, whether from the extended HB model or from the fixed  
$\eta$=0.4 case, implies an age around 5-6~Gyrs, whilst for the solar 
metallicity case the inferred age is $<$3~Gyr
(remembering that the underlying population is a 14~Gyr one in all cases.)

Looking in more detail at the individual models, it is evident  
that if a particular HB simulation has a distribution with a mean mass 
which corresponds to $T_{eff} \sim$9000-9500K, then H$\beta$ is 
maximised, implying spuriously young ages from the integrated SSP spectra 
(confirming the result noted by \citealt{lee2000}).

We created bi-modal HBs by combining models HB1 and HB4 in each set, and
varying the fractions of the red and blue components.  In all cases it was 
found that the change in H$\beta$ strength just scales linearly with the 
change in blue HB fraction.  This is illustrated in Figures~\ref{fig:aegrid} 
and \ref{fig:ssgrid} where the open star symbol represents the 50/50 case
(i.e. 50\% red, 50\% blue HB) which always lies exactly half way between 
model HB1 and model HB4. Therefore inferred ages for an underlying 14 Gyr 
population can lie anywhere between $<$3 and 14 Gyr depending on the precise
details of the HB morphology and the metallicity of the population.  

In general, the measured EWs of the various metal lines are only marginally 
affected by the HB extension, however inferred metallicities can still
change if H$\beta$ increases significantly, since none of the 
diagnostic grids is completely orthogonal.  In practice however, these 
inferred changes in metallicity are likely to be within the observational 
errors in most cases.

\section{Identifying extended blue HB tracers}
\label{sec:ident}

\begin{figure}
\includegraphics[width=80mm,angle=0]{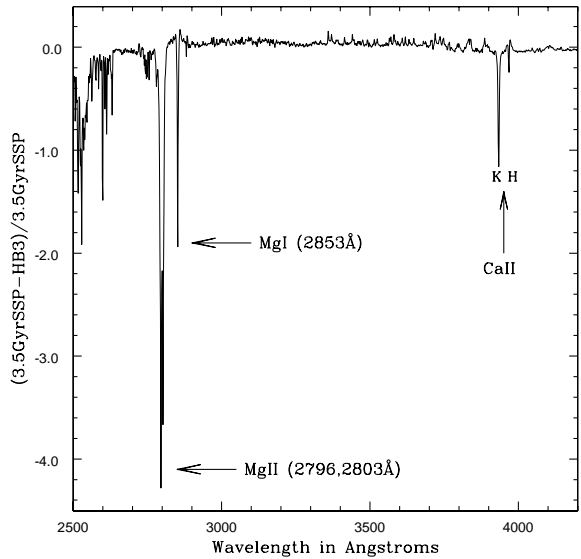}
\caption{Fractional difference spectrum between HB model 3 (14~Gyr plus 
extended blue HB) and a 3.5~Gyr SSP with $\eta$=0.2, from the 
$\alpha$-enhanced, [Fe/H]=$-$0.7 set of models. Spectra were normalised
at 6000\AA. }
\label{fig:diffspec}
\end{figure}

One of the main issues raised by the results presented here is 
whether there are any diagnostic indices which could potentially 
distinguish between an old population with an extended blue HB and a 
simple, single-aged population with an intermediate age (between, say, 
3 and 8~Gyrs) in the integrated spectrum of an unresolved stellar system.
This point has been raised before, but generally in the context of Galactic
globular clusters, which all have sub-solar metal abundances and where 
it is known from isochrone fitting that the true ages are $>$10~Gyr.  
Here we demonstrate quite clearly that any old population, even at solar 
and super-solar metallicities, can have a hot HB if there is sufficient 
mass loss (coupled with a spread in mass loss) on the RGB, and that this
will have the effect of making that population look spuriously young when 
deriving ages from Balmer lines.  
This is not merely an academic point since it is known that 
the compact elliptical galaxy M32, which has near solar metallicity, 
has just such an extended blue HB component, which appears to centre 
around 9000-10000K \citep{brown}, the temperature at which the H$\beta$ 
line is strongest.  It is not clear what fraction of the total population
contributes to this blue HB as this is difficult to quantify from
observational data, but it is known that the spectroscopic age of M32, 
from its Balmer lines, is around 3~Gyr \citep{schi-m32}.

In order to help in the identification of potentially discriminating
features in the spectra, we compared the integrated spectrum of the
HB model 3 case (14~Gyr, with an extended blue HB, maximising H$\beta$) with
a 3.5~Gyr SSP at the same metallicity.  We used the $\alpha$-enhanced, 
[Fe/H]=$-$0.7 set of models for this purpose.  It can be seen from 
Figure~\ref{fig:aegrid} that the H$\beta$ and Fe5406 index strengths are 
almost exactly the same for the 2 scenarios, as indeed are the strengths
of other metal indicators such as Mg$b$ and the various Fe lines (not 
illustrated here).  The flux for the 2 spectra was normalised to unity at 
6000\AA\ and a fractional difference calculated in the sense 
(3.5GyrSSP $-$ HBmodel3)/3.5GyrSSP.  
The resulting fractional difference spectrum is shown in 
Figure~\ref{fig:diffspec} which shows that all the differences between the
2 cases lie at wavelengths shorter than $\sim$4000\AA\ and, in fact, longwards 
of 4500\AA\ this difference spectrum is almost featureless.  The 3 features 
which obviously stand out are the Ca{\sc ii} feature made up of the H and K lines 
(3968.5\AA\ and 3933.7\AA), MgI (2853\AA) and the Mg{\sc ii} doublet (2796\AA\ and 
2803\AA).  It is also clear that the continuum levels shortward of the 
magnesium features are different in the sense that the 14~Gyr population with
extended blue HB has stronger UV continuum than the 3.5~Gyr population.

We note here that we also performed a similar test for the solar 
metallicity, scaled solar models, comparing the spectrum of HB model 
4 with a 3~Gyr SSP at the same metallicity, and found quantitatively 
very similar results to the $\alpha$-enhanced case described above.

\subsection{The Mg{\sc ii} index}
\label {sec:mgII}

\begin{figure}
\includegraphics[width=80mm,angle=0]{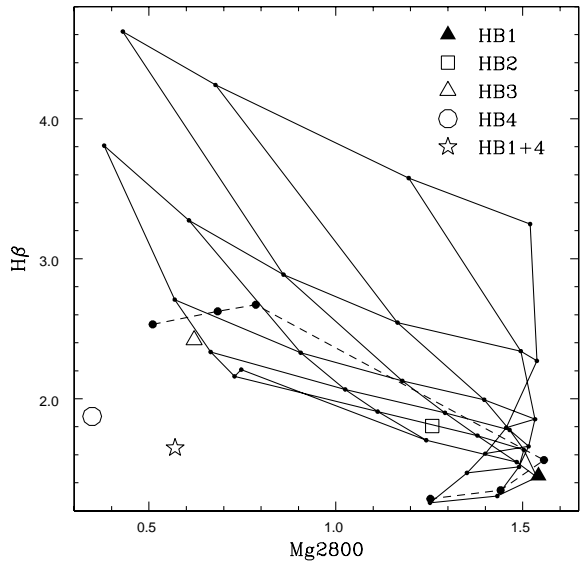}
\caption{H$\beta$ vs. Mg2800 $\alpha$-enhanced grid.  The underlying grid, 
model datapoints and symbols are as for Figure~\ref{fig:aegrid} -- only the
[Fe/H]=$-$0.7 set of models is shown, for clarity.}
\label{fig:mg2800grid}
\end{figure}

At first glance it appears that the Mg{\sc ii} feature around 2800\AA\ is 
the most obvious candidate to investigate as an extended blue HB discriminator.
This index, which is measured as an equivalent width, was defined by 
\citet{fanelli} who were looking for temperature, luminosity, and metallicity 
discriminants for stellar spectra in the mid-UV.  They found Mg{\sc ii} displays 
``no sensitivity to abundance for cool stars and a reversed sensitivity in 
FG dwarfs such that metal-poor stars have stronger Mg{\sc ii} strengths at the 
same temperature than more metal-rich stars''.
We calculated the Mg{\sc ii} equivalent widths from the
integrated spectra of all the models created here, as well as for the 
BaSTI $\eta$=0.2 and $\eta$=0.4 models used for our underlying grids.  The
resulting H$\beta$ vs. Mg{\sc ii} $\alpha$-enhanced diagram is shown in 
Figure~\ref{fig:mg2800grid} (for clarity, only the HB model set at 
[Fe/H]=$-$0.7 is plotted).  This grid clearly shows the behaviour referred
to by \citet{fanelli} as the higher metallicity end of the grid (i.e. towards 
the right of the diagram) folds back on itself, so that there is a large
degree of degeneracy at old ages for all metallicities above 
[Fe/H]$\sim-1.0$, at least for the fixed $\eta$=0.2 case.

More promisingly, there is a strong trend towards lower Mg{\sc ii} values as the
HBs become more extended towards the blue.  Figure~\ref{fig:mg2800grid} 
shows that the 
HB models 3 and 4 lie outside the standard grid, implying a much lower 
metallicity than the true one.  However, because the grid is not very
orthogonal, the implied age is actually close to the true age of 14~Gyr.
Interestingly the Lick index Mg$b$ behaves similarly to the Fe and [MgFe]
indices displayed in Figures~\ref{fig:aegrid} and \ref{fig:ssgrid}, i.e.
it is negligibly affected by extended blue HBs, and so this index
{\it does} give a good indication of the true metallicity.  Hence any 
discrepancy between the implied metallicities from the Mg$b$ and Mg{\sc ii} 
indices for a given stellar population could be due to the presence of
an extended blue HB component.  

There is tentative evidence that Mg{\sc ii} could be used in this way in the UV
data presented in \citet{ponder}.  Their Table 4 includes Mg{\sc ii} data for
several Galactic GCs, including the well known pair M3 (NGC~5272) and 
M13 (NGC~6205), which are known to have almost identical metallicities (around
[Fe/H]=$-$1.5) and similar ages, to within $\sim$1~Gyr.  However their
HB morphologies are quite different, with that of M13 having a very 
extended blue tail.  The \citet{ponder} data show that their Mg{\sc ii} strengths 
are different by 0.26 in the sense that M13 value is lower, which is
qualitatively and quantitatively in reasonable agreement with our 
predictions.

An important caveat to the behaviour of Mg{\sc ii} discussed here is that the 
current version of our population synthesis models does not include 
post-AGB stars, which are expected to emit much of their flux 
at UV wavelengths.  Whilst an order of 
magnitude estimate indicates that their contribution to the continuum
flux of an integrated spectrum is likely to be very small, it would be 
unwise to assume that they would not affect line indices.  However, this
magnesium feature is clearly an intriguing possibility and merits 
further investigation.

\subsection{The Ca{\sc ii} index}
\label {sec:caII}

\begin{figure}
\includegraphics[width=80mm,angle=0]{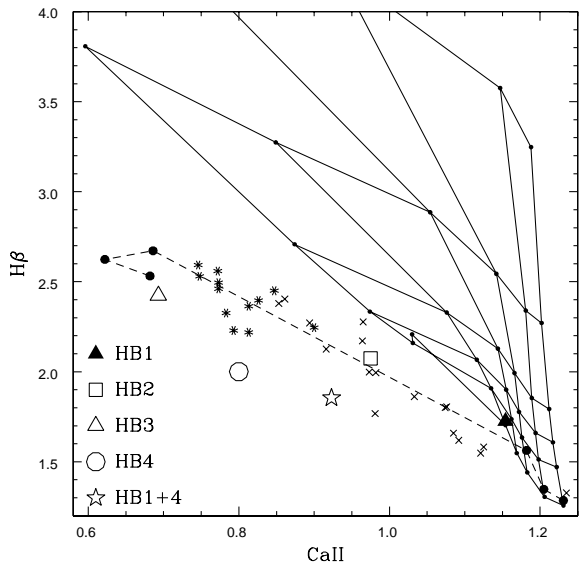}
\caption{H$\beta$ vs. Ca{\sc ii} $\alpha$-enhanced grid.  The underlying grid, 
model datapoints and symbols are as for Figure~\ref{fig:aegrid} -- only the
[Fe/H]=$-$1.3 set of models is shown, for clarity. 
Crosses and asterisks are observational data for Galactic GCs from 
\citet{schi05} -- the asterisks denote clusters with HB type $>$ 0.8 
(i.e. predominantly blue).}
\label{fig:caIIgrid}
\end{figure}

The Ca{\sc ii} index was identified by \citet{rose84} as being highly sensitive
to the presence of A stars in a composite spectrum.  This index is constructed
by dividing the central intensity of the Ca{\sc ii}~H~+~H$\epsilon$ line by the 
central intensity of the Ca{\sc ii}~K line.  
\citet{rose84} states that ``this ratio, which is 
constant in F, G, and K stars, decreases dramatically in 
A and B stars as the Ca{\sc ii} lines weaken and H$\epsilon$ strengthens''.  
Since A stars have effective temperatures in the range 8000K-10000K, which 
encompasses the range of the extended blue HB stars of interest here, 
this index merits a more detailed analysis.  
Hence we calculated Ca{\sc ii}~H~+~H$\epsilon$/Ca{\sc ii}~K from the
integrated spectra of all the models created here, as well as for the 
BaSTI $\eta$=0.2 and $\eta$=0.4 models used for our underlying grids.  The
resulting H$\beta$ vs. Ca{\sc ii} $\alpha$-enhanced grid is shown in 
Figure~\ref{fig:caIIgrid} (for clarity, only the HB model set at 
[Fe/H]=$-$1.3 is plotted).  

The first striking feature of this plot is that the lower metallicity 
14~Gyr SSPs with $\eta$=0.4 (the dashed line) lie outside the $\eta$=0.2 
grid.  Compare this with the H$\beta$ vs. Fe5406 grid in 
Figure~\ref{fig:aegrid}, and the H$\beta$ vs. Mg{\sc ii} grid in 
Figure~\ref{fig:mg2800grid},  where the $\eta$=0.4 models lie completely
within the $\eta$=0.2 grid.  Perhaps more importantly, all the extended 
HB models also lie outside the $\eta$=0.2 grid, except for HB model 1, 
which is the red clump case which has no blue component (and thus behaves 
very similarly to the fixed $\eta$=0.2 case).  This means that the 
combination of H$\beta$ and the Ca{\sc ii} index could potentially 
distinguish between an old population with extended blue HB and an 
intermediate age or young SSP.
Overplotted on Figure~\ref{fig:caIIgrid} are Galactic globular cluster data
from \citet{schi05} for which we have measured the indices directly on the
observed spectra in exactly the same way as for our models.  It is 
interesting to note that these clusters, which all have sub-solar 
metallicity and extended HBs, all lie in the region of the diagram covered 
by our extended HB models.  In fact, closer inspection of individual clusters
reveals that the clusters with the bluest HBs do indeed fall towards the
the left hand side of this diagram, with lower Ca{\sc ii} values.  Clusters
with HB~type~$>$~0.8 (i.e. predominantly blue) have been plotted as asterisks
and it can be seen that they all fall well outside the fixed $\eta$=0.2
`standard' grid.  Looking again at the H$\beta$ vs. Fe5406 grid in 
Figure~\ref{fig:aegrid} it can be seen that these same blue HB clusters 
would all appear to have ages around 6-7~Gyr in that plane.  If their
strong H$\beta$ was really being caused by younger ages rather than blue 
HBs, then these clusters would have a higher Ca{\sc ii} ratio for the same 
H$\beta$ strength and the points would lie further towards the right
in the H$\beta$ vs. Ca{\sc ii} grid. 

An important caveat to note regarding the Ca{\sc ii} index is that it is strongly
affected by velocity dispersion, and also requires very high signal-to-noise
data.  This is because it is measured simply as the ratio of the depths
at the central points of the Ca H and K lines, and is not an equivalent 
width (or magnitude) as most of the other diagnostic indices in general use 
are.  This means it is not an ideal choice for studying elliptical galaxies,
unless their velocity dispersions are known to be low.  However it may be a 
useful indicator for application to extragalactic globular clusters, if 
sufficiently high signal-to-noise spectra can be obtained to measure the
relative depths of the Ca lines accurately.  Further issues relating to the 
Ca{\sc ii} index and its use are discussed in Section~\ref{sec:disc}.

\section{Statistical fluctuations tests}
\subsection{Method}
\label{sec:statfluc}

In order to assess the likely impact of statistical fluctuations on 
integrated spectra, and hence diagnostic line indices, for real stellar 
systems, we simulated 2 cases representing typical Galactic 
globular clusters (GCs).
These 2 cases were based on the 14~Gyr, [Fe/H]=$-1.31$ ($\alpha$-enhanced) 
SSP set of models listed in Table 1. Specifically we used Model 1 and 
Model 3 as the basis for these simulations, which are, respectively, 
the red clump HB case which does not significantly affect H$\beta$, 
and the intermediate extension HB case, which results in the H$\beta$ 
peak seen in Figure~\ref{fig:aegrid}.  For these tests, the whole of the
underlying isochrone up to the TRGB, plus the extended HB, were populated 
with a discrete number of stars to represent a total cluster mass
of $2\times10^5M_\odot$. For individual points along the RGB the mass loss 
was varied to give a spread of masses on the HB and hence a horizontal 
extension, as described in Section~\ref{sec:modelmethod}.  

We also modelled a representative low mass cluster using a 3.5~Gyr, 
[Fe/H]=$-$0.7 SSP as the underlying population, for which we adopted 
a total cluster mass of $2\times10^4M_\odot$.  These parameters were chosen to
simulate a stellar cluster typical of those found in the Magellanic Clouds,
such as those studied in \citet{dias}.  The reasons for this choice will
be discussed further in Section~\ref{sec:disc}.
Observationally, this type of cluster would typically have a red clump HB 
(model 1) which, although extended rather than fixed point, is negligibly 
different from the fixed $\eta$=0.2 case (as demonstrated in 
Section~\ref{sec:hbresults}).  
  
For all 3 of these systems, we ran Monte Carlo simulations, creating 
30 realisations of each case.
To explore the likely maximum mass at which statistical fluctuations might
be significant in stellar clusters, we also created 30 realisations of 
the 14 Gyr extended blue HB case adopting a cluster mass of $2\times10^6M_\odot$.

\begin{figure}
\includegraphics[width=80mm,angle=0]{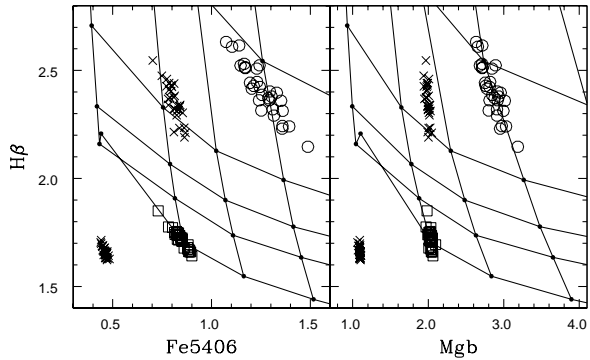}
\caption{Diagnostic grids with results from 30 realisations 
of each of the 3 models described in the text (lines of constant age and
[Fe/H] are as in Figure~\ref{fig:aegrid}): $2\times10^5M_\odot$ GC with HB 
model 1 ({\it open squares}); $2\times10^5M_\odot$ GC with HB model 3 
({\it crosses}); $2\times10^4M_\odot$ open cluster with HB model 1 
({\it open circles}).  Data points in the lower left of each diagram are
30 realisations of the GC with HB model 3 modelled using $2\times10^6M_\odot$, 
which have been shifted arbitrarily for clarity -- as measured they 
are centred on the same location as the $2\times10^5M_\odot$ data.}
\label{fig:sampling}
\end{figure}

\subsection{Results}
\label{sec:sfresults}

\begin{table}
 \centering
 \begin{minipage}{60mm}
  \caption{1$\sigma$ variations in 21 Lick indices, plus H$\delta_{F}$, H$\gamma_{F}$ and Ca{\sc ii}, from the (30) extended blue HB simulations at $2\times10^5M_\odot$.}
  \begin{tabular}{lcc}
  \hline
Index & 1$\sigma$ & Units \\
  \hline
H$\delta_{F}$   &     0.129  &  \AA   \\
H$\gamma_{F}$   &     0.143  &  \AA   \\
CN$_{1}$        &     0.003  &  mag   \\
CN$_{2}$        &     0.003  &  mag   \\
Ca4227          &     0.028  &  \AA   \\
G4300           &     0.134  &  \AA   \\
Fe4383          &     0.091  &  \AA   \\
Ca4455          &     0.028  &  \AA   \\
Fe4531          &     0.069  &  \AA   \\
C$_{2}$4668     &     0.042  &  \AA   \\
H$\beta$        &     0.084  &  \AA   \\
Fe5015          &     0.106  &  \AA   \\
Mg$_{1}$        &     0.003  &  mag   \\
Mg$_{2}$        &     0.004  &  mag   \\
Mg$b$           &     0.024  &  \AA   \\
Fe5270          &     0.054  &  \AA   \\
Fe5335          &     0.057  &  \AA   \\
Fe5406          &     0.035  &  \AA   \\
Fe5709          &     0.032  &  \AA   \\
Fe5782          &     0.017  &  \AA   \\
Na D            &     0.027  &  \AA   \\
TiO$_{1}$       &     0.001  &  mag   \\
TiO$_{2}$       &     0.001  &  mag   \\
Ca{\sc ii}            &     0.015  &  --    \\ 
\hline
\end{tabular}
\end{minipage} 
\label{tab:1sigma}
\end{table}

Results from the statistical fluctuation tests are displayed in 
Figure~\ref{fig:sampling} which shows the H$\beta$ vs. Fe5406 and  
H$\beta$ vs. Mgb diagnostic grids.  For the $2\times10^5M_\odot$ GC simulations
the scatter in H$\beta$ is significant, and is larger for the extended 
blue HB case.  However it is difficult to rigorously quantify the impact 
on implied ages for 3 reasons.  Firstly, for the same degree of scatter
in H$\beta$, the scatter in apparent age depends on the absolute H$\beta$
strength, since H$\beta$ strength does not vary linearly with age.  
For example, a change of $\pm$0.2 in H$\beta$ implies a shift of 
$\sim$1~Gyr for a 3~Gyr SSP, but a shift of $\sim$4~Gyr for a 12~Gyr SSP. 
Secondly,  for the extended blue HB case, the implied age is dominated by the 
large offset towards younger ages caused by the blue HB itself, i.e. here the 
underlying 14~Gyr SSP appears to have an age $\sim$5~Gyr, even without the
statistical fluctuations (see Figure~\ref{fig:aegrid}).  
Thirdly, the magnitude of the  scatter in apparent age caused by the 
scatter in H$\beta$ also depends on which diagnostic grid 
is used and what the properties of the underlying population are.  
Figure~\ref{fig:sampling} illustrates this last point.  In the 
left hand panel (i.e. the H$\beta$ vs. Fe5406 grid) it can be seen that the 
Fe5406 index also has a scatter of around $\pm$0.1, in addition to the 
scatter in H$\beta$.  For the extended 
blue HB case this induces an uncertainty on both the implied age and
metallicity of the population.  However for the red HB case (model 1), 
the magnitude of the fluctuations for these 2 indices conspires to scatter
the points along a line of constant age, implying no uncertainty on the age.
The right hand panel of Figure~\ref{fig:sampling} demonstrates that not
all metallicity indicators behave similarly however -- in this case, the Mgb
index shows only negligible fluctuations and all the scatter is seen in
the H$\beta$ index, implying an uncertainty of $\pm$2-3~Gyr for the same 
red HB simulations.

As a guide to the relative significance of fluctuations in various 
diagnostic indices, 1$\sigma$ variations from the 30 simulations of the 
extended blue HB case at $2\times10^5M_\odot$ are tabulated in Table~3 for 
all 21 Lick indices, plus H$\delta_{F}$, H$\gamma_{F}$ and Ca{\sc ii}.  

As might be expected, fluctuations in H$\beta$ are larger for the 
$2\times10^4M_\odot$ case with a scatter of around $\pm$0.25, implying an
uncertainty of up to 2~Gyrs on the modelled 3.5~Gyr SSP.  It should be noted 
that, in this case, most of the fluctuations are coming from discrete 
sampling in the upper portion of the RGB, since the HB consists of a very 
compact red clump with very little extension.  

Results from the $2\times10^6M_\odot$ simulations are also plotted in 
Figure~\ref{fig:sampling} but have been arbitrarily shifted to the lower
left hand corner of the diagrams for clarity.  It can be seen that, even with
this total mass, there are still fluctuations in the H$\beta$ index at the
level of $\sim$0.05 in EW, corresponding to an uncertainty in age of 
0.5-1~Gyr for old SSPs.  Various of the metal lines also display low
level fluctuations implying a minimum uncertainty of around 0.05~dex in
[Fe/H].  However, for these $2\times10^6M_\odot$ simulations all indices only
fluctuate at a level that is likely to be within the measurement errors for 
real data.

\section{Summary and discussion}
\label{sec:disc}

To summarise, we have created integrated spectra for 16 SSPs, 4 each
at 4 different metallicities, all with an underlying age of 14~Gyr, with a 
range of extended HB morphologies.  This was done by varying the mass loss
prescription for each individual SSP in 2 ways, firstly setting a mean mass, 
$<M>$, for stars coming on to the ZAHB, and then adding a spread in mass loss, 
$\sigma_{M}$.  We find that the H$\beta$ strength for each SSP depends 
on the exact temperature distribution of stars along the HB, which in turn
depends on the exact details of the mass loss prescription coupled with 
the metallicity of the population in question.  For any of the modelled SSPs 
with any amount of blue HB extension, H$\beta$ is increased relative 
to the fixed $\eta$=0.2 case (i.e. the `standard' models), 
implying younger ages than the actual SSP age of 14~Gyr.  
In the worst-case scenario modelled here, a solar
metallicity 14~Gyr population with a blue HB which has a peak in its 
distribution of stars around 9000K, has an implied age of around 2~Gyr from
the strength of the H$\beta$ line.

Our preliminary investigation to identify spectral features which might be
capable of breaking the degeneracy between an old SSP with extended
blue HB, and a truly intermediate age or young SSP, indicates that the Ca{\sc ii}
index defined by \cite{rose84} is a very promising candidate, at least for
populations with low velocity dispersions, such as extra-galactic globular
clusters (but see caveat in the following paragraph).  
There is also tentative evidence that the Mg{\sc ii} index, in combination with 
Mg$b$, could also be a very useful tracer of hot blue HBs.  
There are also indications that the UV continuum shortward of the Mg{\sc ii} 
feature could also potentially be a useful additional tool, although more
work is needed to model this part of the spectrum in the required detail.  
However, it is interesting
to note that the extreme blue HB models (i.e. models 3 and 4 presented here) 
have very strong UV continuum flux, as demonstrated by 
Figure~\ref{fig:diffspec}, even at solar metallicity.  This supports the
idea that extreme blue HBs can be a significant contributor to the UV upturn 
identified in old elliptical galaxies (see review by \citealt{oconnell}).

An important caveat to all the tests and results presented here is that 
all our models have been created using SSPs, i.e. single age, single 
metallicity systems.  However another scenario which could explain
strong H$\beta$ lines in a predominantly old population would be the presence
of a small fraction of very young stars.  Although this is not likely to
be an issue for globular clusters (Galactic or extra-galactic) it is a
possibility which is hard to completely rule out for elliptical galaxies.
Preliminary tests with our population synthesis code indicate that adding 
even a very small percentage ($<$1\%) of a young SSP (300~Myr or less) to 
a 14~Gyr SSP would significantly strengthen H$\beta$ and imply an intermediate
age, around 5-6~Gyr (see also \citealt{serra}).  
In fact, this scenario of a `frosting' of young stars in an otherwise
old stellar population has been explored by \citet{smith} for 
(apparently) quiescent galaxies in the Shapley supercluster.  However, 
\citet{smith} use the Ca{\sc ii} index as a tracer of hot stars and, as we have
shown here, this index also traces hot HB stars in an SSP.  So far we have 
not identified any completely unambiguous tracers that can distinguish between
the 3 scenarios; 1) a small fraction of hot young stars in an otherwise old 
population, 2) an old SSP with a hot HB, 3) an intermediate age SSP.  It seems
likely that no single tracer will be able to disentangle these 3 cases
and a combination of well understood spectral indices and colours 
may well be needed -- this is the focus of our ongoing investigation.


Another factor potentially linked to extended blue HBs is the presence
of a stellar subpopulation with a high helium fraction.  \citet{lee05} 
showed that the same level of helium enrichment required to reproduce the 
bluer main sequence in the massive GC $\omega$ Centauri would also naturally
produce the extreme blue HB stars seen in that cluster.  
Very recent observational results have provided some evidence for varying 
helium fractions within individual GCs and there is at least 
circumstantial evidence linking helium enriched subpopulations to
extended blue HBs (see \citealt{brag2010} and references therein).
The theoretical support for the HB morphology--enhanced-helium connection is 
that stars born with a higher helium abundance will display a lower mass at the 
turn off for a given cluster age.  A lower mass at the TO will favour 
lower mass -- hence hotter and bluer -- HB stars.

We were able to test the potential impact of a higher helium abundance on
our present work by utilising the helium-enhanced isochrones in the BaSTI 
database in combination with a grid of helium-enhanced spectra, which
have been produced for a subsequent paper (Coelho, Salaris \& Percival, 2011, 
in preparation).  We chose helium-enhancement at the level $Y=0.3$, for the
[Fe/H]=$-0.7$ isochrone, which is a reasonable enhancement over the 
standard cosmological value, given current observational constraints 
(e.g. \citealt{brag2010}) and our new spectra incorporate the same Fe and He
abundance ratios as the isochrones.  
As a preliminary test, we measured the strength of spectral features in 
individual spectra for both the standard helium and enhanced helium versions, 
matching the spectra in $T_{eff}$ and log$g$.  We found that for all the 
diagnostic lines considered in this paper, the line strengths are 
practically indistinguishable for the two cases, with any differences
being less than 1 percent (i.e. within the measurement errors).  We also
created an integrated spectrum for the [Fe/H]=$-0.7$, 14~Gyr SSP using the
helium-enhanced isochrone in combination with the helium-enhanced spectra.  
Again we 
found that the strengths of all the indices under scrutiny here are 
negligibly different from the standard helium case.  We note here that this 
result is consistent with that of \citet{girardi2007} who find that a similar
level of helium enhancement has a negligible effect on broadband colours.

The important point to stress here is that, whilst enhanced helium is a
likely mechanism for producing extended blue HBs, it in no way impacts
on the work presented in this study.  Our diagnostic models and 
tests are not intended
to predict, $a~priori$, the existence of hot HB stars, but rather to 
find diagnostics which can distinguish their presence in a stellar 
population from other hot components, such as a young sub-population.  
Whether a blue HB morphology arises from a large mass loss along the 
RGB in stars with `normal' helium, or from a more moderate mass loss from 
stars with higher helium, the resulting integrated spectrum is largely 
unaffected -- the only parameter that matters here is the temperature 
range of stars on the HB, irrespective of how they have been produced.
In fact, the results of our enhanced helium tests, described above, show that
enhanced helium has absolutely no impact on the diagnostic indices discussed 
here, such as H$\beta$,  Mg{\sc ii} and  Ca{\sc ii}, nor on the results of the 
statistical fluctuations tests discussed below.

Finally, the results of the tests presented here in 
Section~\ref{sec:statfluc} demonstrate that several key diagnostic line 
indices are significantly affected by statistical fluctuations, even in 
average to large mass GCs.  Equally problematic is the fact that not all
indices fluctuate at the same level -- some are strongly affected, 
including all the Balmer lines, whilst others are only negligibly 
affected, such as Ca{\sc ii} (see Table 3).  This is potentially a significant
source of uncertainty if full-SED fitting methods are used to derive ages,
metallicities and/or star formation histories from integrated spectra, since
different features in a spectrum can be giving conflicting best fit 
parameters.  The problem is illustrated in \citet{dias} who present integrated
spectra for 14 Magellanic Cloud stellar clusters for which they derive best
fit ages and metallicities using 2 different fitting codes, {\sc starlight} 
\citep{cidf} and $ULySS$ \citep{koleva}, and 3 different sets of SSP models,
from \citet{bc03}, \citet{pegase} (PEGASE-HR) and \citet{vaz2010}.
Half of the clusters studied have masses in the range $1-2.5\times10^4M_\odot$,
which corresponds to the low mass cluster modelled here, in 
Section~\ref{sec:statfluc}.  For several clusters, \citet{dias} find that
the results from the different fitting routines and models are completely
discrepant.  As an example, cluster HW1 yields best-fit ages of 3.2, 5.8,
7.9, 9.0, 9.4 and 10~Gyr from the 6 different combinations of
SSPs models and SED-fitting routines, whilst its actual age is known to be
around 6~Gyr from isochrone fitting to the CMD.  Other clusters are even more
unconstrained, yielding ages ranging from $<$1~Gyr to 10~Gyr for the same 
cluster.  Some of these variations in best fit parameters are likely 
to be due to systematic uncertainties in stellar parameters inherent 
within the SSP models themselves (see \citealt{smp}), but statistical
fluctuations in the observed spectra are almost certainly contributing 
to the problem.  Hence we urge caution if these types of stellar
clusters are to be used as empirical calibrating objects for various 
aspects of SPS models.

\section*{Acknowledgments}
We thank PAJ and DXC for useful discussions and comments on an early 
version of this paper.  
We also thank editorial staff at MNRAS for their handling of the manuscript.

SMP acknowledges financial support from the Science 
\& Technology Facilities Council (STFC) through a Postdoctoral Research 
Fellowship.

\bsp


\begin{thebibliography}{}

\bibitem[\protect\citeauthoryear{{Bragaglia}, {Carretta}, {Gratton}, {D'Orazi},
  {Cassisi} \& {Lucatello}}{{Bragaglia} et~al.}{2010}]{brag2010}
{Bragaglia} A.,  {Carretta} E.,  {Gratton} R.,  {D'Orazi} V.,  {Cassisi} S.,
  {Lucatello} S.,  2010, A\&A in press, ArXiv:1005.2659

\bibitem[\protect\citeauthoryear{{Brown} \& {Ferguson}}{{Brown} \&
  {Ferguson}}{2003}]{brown}
{Brown} T.~M.,  {Ferguson} H.~C.,  2003, in {G.~Piotto, G.~Meylan,
  S.~G.~Djorgovski, \& M.~Riello} ed., New Horizons in Globular Cluster
  Astronomy Vol.~296 of Astronomical Society of the Pacific Conference Series,
  {Using M32 to Study Rapid Phases of Stellar Evolution}.
p.~199

\bibitem[\protect\citeauthoryear{{Bruzual} \& {Charlot}}{{Bruzual} \&
  {Charlot}}{2003}]{bc03}
{Bruzual} G.,  {Charlot} S.,  2003, MNRAS, 344, 1000

\bibitem[\protect\citeauthoryear{{Cid Fernandes}, {Mateus}, {Sodr{\'e}},
  {Stasi{\'n}ska} \& {Gomes}}{{Cid Fernandes} et~al.}{2005}]{cidf}
{Cid Fernandes} R.,  {Mateus} A.,  {Sodr{\'e}} L.,  {Stasi{\'n}ska} G.,
  {Gomes} J.~M.,  2005, MNRAS, 358, 363

\bibitem[\protect\citeauthoryear{{Coelho}, {Bruzual}, {Charlot}, {Weiss},
  {Barbuy} \& {Ferguson}}{{Coelho} et~al.}{2007}]{coelho07}
{Coelho} P.,  {Bruzual} G.,  {Charlot} S.,  {Weiss} A.,  {Barbuy} B.,
  {Ferguson} J.~W.,  2007, MNRAS, 382, 498

\bibitem[\protect\citeauthoryear{{Conroy}, {Gunn} \& {White}}{{Conroy}
  et~al.}{2009}]{conroy09}
{Conroy} C.,  {Gunn} J.~E.,    {White} M.,  2009, ApJ, 699, 486

\bibitem[\protect\citeauthoryear{{Cordier}, {Pietrinferni}, {Cassisi} \&
  {Salaris}}{{Cordier} et~al.}{2007}]{basti3}
{Cordier} D.,  {Pietrinferni} A.,  {Cassisi} S.,    {Salaris} M.,  2007, AJ,
  133, 468

\bibitem[\protect\citeauthoryear{{Dias}, {Coelho}, {Kerber}, {Barbuy} \&
  {Idiart}}{{Dias} et~al.}{2010}]{dias}
{Dias} B.,  {Coelho} P.,  {Kerber} L.,  {Barbuy} B.,    {Idiart} T.,  2010,
  A\&A in press, ArXiv:1002.4301

\bibitem[\protect\citeauthoryear{{Fanelli}, {O'Connell}, {Burstein} \&
  {Wu}}{{Fanelli} et~al.}{1990}]{fanelli}
{Fanelli} M.~N.,  {O'Connell} R.~W.,  {Burstein} D.,    {Wu} C.,  1990, ApJ,
  364, 272

\bibitem[\protect\citeauthoryear{{Girardi}, {Castelli}, {Bertelli} \&
  {Nasi}}{{Girardi} et~al.}{2007}]{girardi2007}
{Girardi} L.,  {Castelli} F.,  {Bertelli} G.,    {Nasi} E.,  2007, A\&A, 468,
  657

\bibitem[\protect\citeauthoryear{{Koleva}, {Prugniel}, {Bouchard} \&
  {Wu}}{{Koleva} et~al.}{2009}]{koleva}
{Koleva} M.,  {Prugniel} P.,  {Bouchard} A.,    {Wu} Y.,  2009, A\&A, 501, 1269

\bibitem[\protect\citeauthoryear{{Kotulla}, {Fritze}, {Weilbacher} \&
  {Anders}}{{Kotulla} et~al.}{2009}]{galev09}
{Kotulla} R.,  {Fritze} U.,  {Weilbacher} P.,    {Anders} P.,  2009, MNRAS,
  396, 462

\bibitem[\protect\citeauthoryear{{Le Borgne}, {Rocca-Volmerange}, {Prugniel},
  {Lan{\c c}on}, {Fioc} \& {Soubiran}}{{Le Borgne} et~al.}{2004}]{pegase}
{Le Borgne} D.,  {Rocca-Volmerange} B.,  {Prugniel} P.,  {Lan{\c c}on} A.,
  {Fioc} M.,    {Soubiran} C.,  2004, A\&A, 425, 881

\bibitem[\protect\citeauthoryear{{Lee}, {Lee} \& {Gibson}}{{Lee}
  et~al.}{2002}]{lee2002}
{Lee} H.,  {Lee} Y.,    {Gibson} B.~K.,  2002, AJ, 124, 2664

\bibitem[\protect\citeauthoryear{{Lee}, {Yoon} \& {Lee}}{{Lee}
  et~al.}{2000}]{lee2000}
{Lee} H.,  {Yoon} S.,    {Lee} Y.,  2000, AJ, 120, 998

\bibitem[\protect\citeauthoryear{{Lee}, {Demarque} \& {Zinn}}{{Lee}
  et~al.}{1994}]{lee94}
{Lee} Y.,  {Demarque} P.,    {Zinn} R.,  1994, ApJ, 423, 248

\bibitem[\protect\citeauthoryear{{Lee}, {Joo}, {Han}, {Chung}, {Ree}, {Sohn},
  {Kim}, {Yoon}, {Yi} \& {Demarque}}{{Lee} et~al.}{2005}]{lee05}
{Lee} Y.,  {Joo} S.,  {Han} S.,  {Chung} C.,  {Ree} C.~H.,  {Sohn} Y.,  {Kim}
  Y.,  {Yoon} S.,  {Yi} S.~K.,    {Demarque} P.,  2005, ApJ, 621, L57

\bibitem[\protect\citeauthoryear{{Maraston}}{{Maraston}}{2005}]{maraston05}
{Maraston} C.,  2005, MNRAS, 362, 799

\bibitem[\protect\citeauthoryear{{Munari}, {Sordo}, {Castelli} \&
  {Zwitter}}{{Munari} et~al.}{2005}]{munari}
{Munari} U.,  {Sordo} R.,  {Castelli} F.,    {Zwitter} T.,  2005, A\&A, 442,
  1127

\bibitem[\protect\citeauthoryear{{O'Connell}}{{O'Connell}}{1999}]{oconnell}
{O'Connell} R.~W.,  1999, ARA\&A, 37, 603

\bibitem[\protect\citeauthoryear{{Ocvirk}}{{Ocvirk}}{2010}]{ocvirk}
{Ocvirk} P.,  2010, ApJ, 709, 88

\bibitem[\protect\citeauthoryear{{Percival} \& {Salaris}}{{Percival} \&
  {Salaris}}{2009}]{smp}
{Percival} S.~M.,  {Salaris} M.,  2009, ApJ, 703, 1123

\bibitem[\protect\citeauthoryear{{Percival}, {Salaris}, {Cassisi} \&
  {Pietrinferni}}{{Percival} et~al.}{2009}]{basti4}
{Percival} S.~M.,  {Salaris} M.,  {Cassisi} S.,    {Pietrinferni} A.,  2009,
  ApJ, 690, 427

\bibitem[\protect\citeauthoryear{{Pietrinferni}, {Cassisi}, {Salaris} \&
  {Castelli}}{{Pietrinferni} et~al.}{2004}]{basti1}
{Pietrinferni} A.,  {Cassisi} S.,  {Salaris} M.,    {Castelli} F.,  2004, ApJ,
  612, 168

\bibitem[\protect\citeauthoryear{{Pietrinferni}, {Cassisi}, {Salaris} \&
  {Castelli}}{{Pietrinferni} et~al.}{2006}]{basti2}
{Pietrinferni} A.,  {Cassisi} S.,  {Salaris} M.,    {Castelli} F.,  2006, ApJ,
  642, 797

\bibitem[\protect\citeauthoryear{{Ponder}, {Burstein}, {O'Connell}, {Rose},
  {Frogel}, {Wu}, {Crenshaw}, {Rieke} \& {Tripicco}}{{Ponder}
  et~al.}{1998}]{ponder}
{Ponder} J.~M.,  {Burstein} D.,  {O'Connell} R.~W.,  {Rose} J.~A.,  {Frogel}
  J.~A.,  {Wu} C.,  {Crenshaw} D.~M.,  {Rieke} M.~J.,    {Tripicco} M.,  1998,
  AJ, 116, 2297

\bibitem[\protect\citeauthoryear{{Reimers}}{{Reimers}}{1975}]{reimers}
{Reimers} D.,  1975, Memoires of the Societe Royale des Sciences de Liege, 8,
  369

\bibitem[\protect\citeauthoryear{{Rose}}{{Rose}}{1984}]{rose84}
{Rose} J.~A.,  1984, AJ, 89, 1238

\bibitem[\protect\citeauthoryear{{Schiavon}, {Caldwell} \& {Rose}}{{Schiavon}
  et~al.}{2004}]{schi-m32}
{Schiavon} R.~P.,  {Caldwell} N.,    {Rose} J.~A.,  2004, AJ, 127, 1513

\bibitem[\protect\citeauthoryear{{Schiavon}, {Rose}, {Courteau} \&
  {MacArthur}}{{Schiavon} et~al.}{2004}]{schi04}
{Schiavon} R.~P.,  {Rose} J.~A.,  {Courteau} S.,    {MacArthur} L.~A.,  2004,
  ApJ, 608, L33

\bibitem[\protect\citeauthoryear{{Schiavon}, {Rose}, {Courteau} \&
  {MacArthur}}{{Schiavon} et~al.}{2005}]{schi05}
{Schiavon} R.~P.,  {Rose} J.~A.,  {Courteau} S.,    {MacArthur} L.~A.,  2005,
  ApJS, 160, 163

\bibitem[\protect\citeauthoryear{{Serra} \& {Trager}}{{Serra} \&
  {Trager}}{2007}]{serra}
{Serra} P.,  {Trager} S.~C.,  2007, MNRAS, 374, 769

\bibitem[\protect\citeauthoryear{{Smith}, {Lucey} \& {Hudson}}{{Smith}
  et~al.}{2009}]{smith}
{Smith} R.~J.,  {Lucey} J.~R.,    {Hudson} M.~J.,  2009, MNRAS, 400, 1690

\bibitem[\protect\citeauthoryear{{Trager}, {Worthey}, {Faber}, {Burstein} \&
  {Gonzalez}}{{Trager} et~al.}{1998}]{trager98}
{Trager} S.~C.,  {Worthey} G.,  {Faber} S.~M.,  {Burstein} D.,    {Gonzalez}
  J.~J.,  1998, ApJS, 116, 1

\bibitem[\protect\citeauthoryear{{Vazdekis}, {S{\'a}nchez-Bl{\'a}zquez},
  {Falc{\'o}n-Barroso}, {Cenarro}, {Beasley}, {Cardiel}, {Gorgas} \&
  {Peletier}}{{Vazdekis} et~al.}{2010}]{vaz2010}
{Vazdekis} A.,  {S{\'a}nchez-Bl{\'a}zquez} P.,  {Falc{\'o}n-Barroso} J.,
  {Cenarro} A.~J.,  {Beasley} M.~A.,  {Cardiel} N.,  {Gorgas} J.,    {Peletier}
  R.~F.,  2010, MNRAS, 404, 1639

\bibitem[\protect\citeauthoryear{{V{\'a}zquez} \& {Leitherer}}{{V{\'a}zquez} \&
  {Leitherer}}{2005}]{starb99}
{V{\'a}zquez} G.~A.,  {Leitherer} C.,  2005, ApJ, 621, 695

\end{thebibliography}

\label{lastpage}

\end{document}